\documentclass[10pt, final]{IEEEtran}

\usepackage{amsthm}

\usepackage{amssymb}
\usepackage{amsmath}
\usepackage{bbm}
\usepackage[ruled]{algorithm2e}
\usepackage{mathrsfs}
\usepackage{cite}
\usepackage{CJKutf8}
\usepackage{bm,epsfig,amsthm,url}
\usepackage{indentfirst}
\usepackage{amsfonts}
\usepackage{epstopdf}
\usepackage{cases}
\usepackage[caption=false,font=scriptsize]{subfig}
\usepackage{xcolor}
\usepackage{mathtools}

\makeatletter
\renewcommand{\maketag@@@}[1]{\hbox{\m@th\normalsize\normalfont#1}}%
\makeatother
\usepackage{hyperref}
\usepackage{stfloats}

\newtheoremstyle{mystyle}{}{}{}{}{}{: }{0pt}{\indent \it{\thmname{#1}\thmnumber{ #2}\thmnote{#3}}}
\theoremstyle{mystyle}

\usepackage{graphicx}
\usepackage{url}

\usepackage{tabularx,booktabs}
\newcolumntype{C}{>{\centering\arraybackslash}X} 
\usepackage{lipsum}

\usepackage{makecell} 

\usepackage{graphicx}
\usepackage{color}

\begin{document}

\title{ \fontsize{22pt}{26pt}\selectfont  Intelligent Reflecting Surfaces for THz Communications: Fundamentals, Key Solutions, and System Prototyping}

\author{{Qingqing Wu, Yanze Zhu, Qiaoyan Peng,  Wanming Hao, Yanzhao Hou, Fengyuan Yang, Wencai Yan, Guoning Wang, Wen Chen, and Chi Qiu}
	\thanks{Q. Wu, Y. Zhu, and W. Chen are with the School of Information Science and Electronic Engineering, Shanghai Jiao Tong University,  China (email: \{qingqingwu, yanzezhu, wenchen\}@sjtu.edu.cn). 
    Q. Peng is with the State Key Laboratory of Internet of Things for Smart City, University of Macau, Macao 999078, China, and also with the Department of Electronic Engineering, Shanghai Jiao Tong University, Shanghai, 200240, China (email: qiaoyan.peng@connect.um.edu.mo). Y. Hou is with Beijing University of Posts and Telecommunications (email: houyanzhao@bupt.edu.cn); W. Hao is with Zhengzhou University (email: iewmhao@zzu.edu.cn); F. Yang is with the School of Microelectronics, Shanghai University, China (email: yang\_fengyuan@shu.edu.cn); W. Yan is with the College of Information Science and Engineering, Henan University of Technology, Zhengzhou 450001, China (email: yanwencai@haut.edu.cn); G. Wang is with the School of Computer Science, Shanghai Jiao Tong University, Shanghai, China (email: wguoning@bupt.edu.cn). Chi Qiu is with the School of Computer Science, Hubei University, Wuhan 430062, China (email: chiqiu@hubu.edu.cn). (Corresponding author: Yanzhao Hou).}
}

\maketitle

\begin{abstract} 
Intelligent reflecting surfaces (IRSs) have emerged as a  cost-effective technology for terahertz (THz) communications by enabling programmable control of the wireless environment. This paper provides a comprehensive overview of IRSs-aided THz communications, covering hardware designs, advanced signal processing techniques, and practical deployment strategies. It first examines key THz reconfigurable metasurface architectures, including electronic, optical, phase-change material, and micro-electromechanical systems (MEMS)-based implementations, highlighting their reconfiguration mechanisms and challenges. Then, fundamental effects including near field and beam squint in wideband THz systems are analyzed, along with their impacts on system performance. The paper further explores conventional and beam-squint-assisted channel estimation methods, innovative beam management strategies, and deployment considerations across large- and small-scale scenarios. Practical experiments at 220 gigahertz (GHz) validate the effectiveness of IRS in improving signal strength and communication reliability for both single-user and multi-user setups.  


\end{abstract}


\section{Introduction}

Terahertz (THz) communications, operating in the 0.1–10 THz band, opens up unprecedented opportunities for ultra-high data rate and low-latency applications, making it a key enabler for the sixth-generation (6G) and beyond wireless systems. Despite its great potential, THz communication is hindered by severe propagation losses, limited coverage range,  etc. Existing approaches adopting traditional active antennas and radio frequency (RF) components, such as massive multiple-input multiple-output (MIMO) and extremely large-scale MIMO architectures, can enhance capacity but face critical limitations in practical systems, including high hardware complexity, prohibitive cost, and poor energy efficiency.



To address these limitations, intelligent reflecting surfaces (IRSs) have emerged as a promising paradigm for reconfigurable wireless environments \cite{wu2024intelligent,chen_iot}. By smartly controlling the reflection of THz waves, IRS consisting of massive passive reflecting elements enables passive beamforming and spatial signal shaping at low power and cost. 
Recently, IRS potentials for enhancing THz communications have been found in various application scenarios. For example, in enhanced mobile broadband (eMBB) use cases, IRS  can significantly enhance signal coverage and link capacity, ensuring the high-speed connectivity required for seamless immersive experiences. Furthermore, in future network deployments, IRS offers a promising means to improve both spectral and energy efficiency by steering THz signals toward intended users. Other applications include satellite communications, where IRS assists in overcoming line-of-sight (LoS) challenges, and radar systems, benefiting from improved detection capabilities. 



However, several challenges must be addressed to fully realize the potential of IRS-assisted THz communications. One significant challenge lies in the hardware implementation of IRS at THz frequencies. The intricate nature of THz signal propagation necessitates highly precise and efficient reflecting elements that can dynamically adapt to varying channel conditions. Current materials and fabrication techniques may not meet the stringent requirements for low-loss, high-frequency performance, leading to significant signal degradation. Furthermore, the cost and complexity of manufacturing such advanced materials pose substantial barriers to widespread adoption. On the other hand, the signal processing and networking aspects of IRSs in THz systems present another layer of complexity. The high frequency of THz signals indicates that channel estimation and feedback mechanisms must be exceptionally rapid and accurate to maintain effective communication. Traditional algorithms may not suffice, and new approaches must be developed to handle the unique characteristics of THz channels, including high path loss and sensitivity to environmental factors. These challenges underscore the need for a comprehensive investigation of the current state-of-the-art in IRS technology, specifically tailored for THz applications.

This article provides a holistic investigation  for IRS-assisted THz communications, spanning from theoretical analysis and design to practical system prototyping. First, representative practical THz metasurface implementations are briefly reviewed to highlight their distinct characteristics as compared to conventional sub-6 gigahertz (GHz) counterparts, covering electronic, optical, phase-change, and micro-electromechanical systems (MEMS) based reconfiguration mechanisms.
Then, key THz-specific phenomena such as near-far field propagation and beam squint are systematically analyzed, and their joint impact on beam performance is quantified. Theoretical performance limits are further explored, along with an overview on advanced channel estimation strategies, beam squint mitigation techniques, and deployment solutions. 
Finally, a proof-of-concept THz communication prototype is implemented and evaluated in both single- and multi-user scenarios, confirming the effectiveness of a THz-IRS in enhancing signal quality and supporting dynamic beam steering. This work bridges theory and practice by offering new insights into ultra-high frequency IRS and establishing  concrete guidelines for the design and implementation of IRS-aided THz communications in future wireless networks.





\section{Hardware Designs of THz Metasurfaces}

\begin{figure*}
	\centering
	\includegraphics[width=0.7\textwidth]{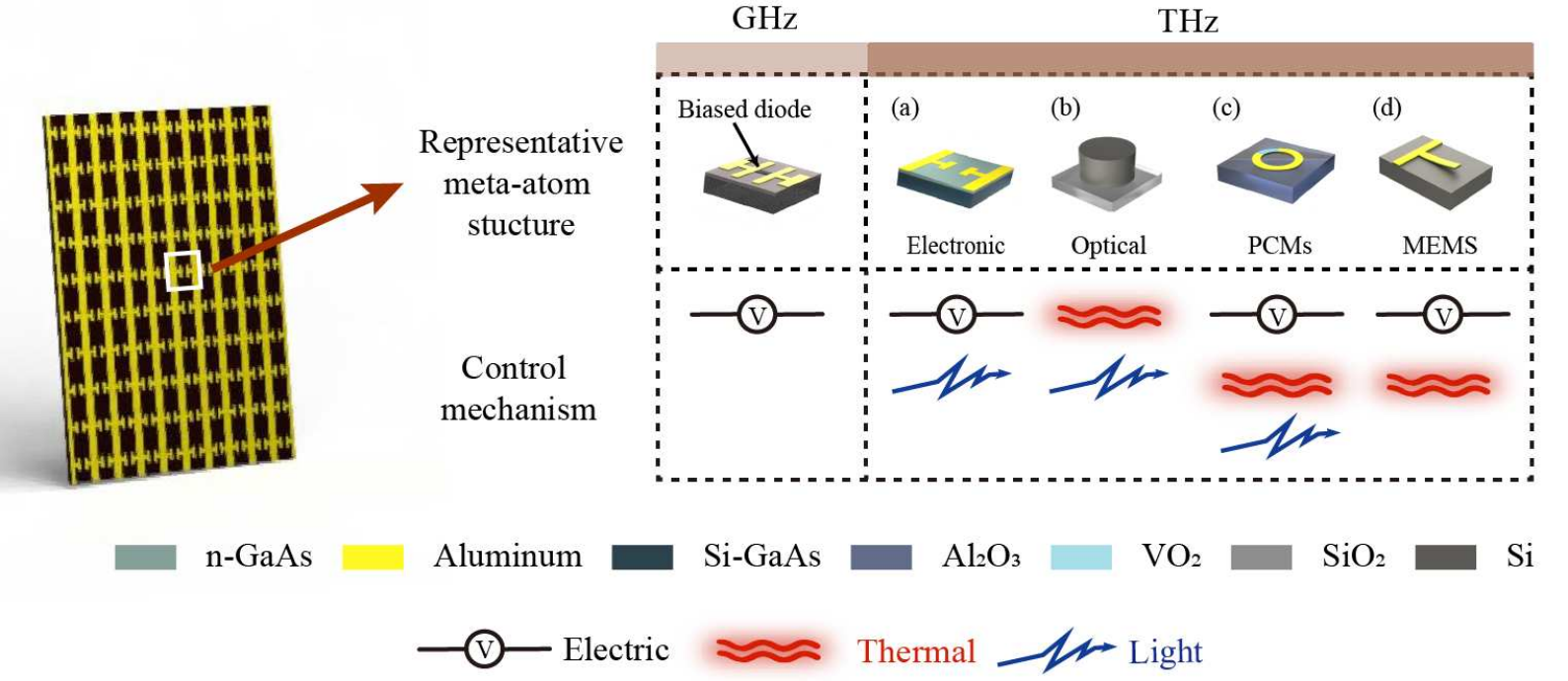}\vspace{-0.3em}\\
	\caption{Illustrations of the THz reconfigurable metasurface employing distinct tuning strategies: a) Electronic approaches using Schottky gate structure; b) Optically driven reconfigurable metasurfaces leveraging semiconductor photoconductivity; c) Phase-change material-enabled metasurfaces (e.g., vanadium dioxide) achieving non-volatile state switching; d) MEMS-based reconfigurable architectures utilizing mechanical actuation for precision structural control.}\label{schematic_illustrations}\vspace{-0.3em}
\end{figure*}

The design of high-frequency reflecting/transmitting metasurfaces serves as the foundation for enabling efficient  channel reconfiguration in THz wireless systems. 
Conventional tunable elements like PIN diodes—widely adopted in sub-6 GHz and millimeter-wave reconfigurable metasurfaces \cite{wu2024intelligent}—struggle at THz frequencies due to inherent physical limitations \cite{switch}. Their switching speeds, constrained by carrier mobility and parasitic capacitance, fall short of the agility required at THz scales ($>$0.1 THz), while metallic interconnects suffer severe insertion losses due to skin-effect dominance at reduced wavelengths. Furthermore, the miniaturization demands of THz meta-atoms push the boundaries of semiconductor processes, introducing quantum tunneling risks that degrade reconfigurability. These challenges necessitate a shift away from scaled-down microwave designs toward native THz metasurfaces—integrating THz-optimized materials and quantum-confined architectures to enable low-loss, high-speed dynamic control.
Emerging solutions span a variety of tuning mechanisms, including electronic and optical approaches, as well as innovative material platforms such as phase-change materials, and mechanically actuated reconfiguration using MEMS. Fig. \ref{schematic_illustrations} illustrates representative meta-atom designs that enable dynamic reconfigurability in THz reconfigurable metasurfaces, which are further elaborated below.
 

\subsubsection{Electronic Approaches}

Traditional active components such as diodes and varactors, though effective in microwave-frequency reconfigurable metasurfaces, encounter significant limitations in THz systems due to their restricted cutoff frequencies. To address these challenges, three advanced electronic tuning strategies have emerged for THz reconfigurable metasurfaces. Specifically, complementary metal oxide semiconductor (CMOS) transistors with optimized structures have been proposed to mitigate parasitic effects at THz frequencies. For example, recent progress in CMOS-based solutions has been reported in \cite{venkatesh2020high}, where a 65 nm CMOS-processed $2 \times 2$ chip array was fabricated, integrating 576 independently addressable meta-atoms. Operating at 0.3 THz, this platform achieves 8-bit digital programming at GHz-scale speeds with dynamic amplitude modulation and phase control supporting functions such as multi-beamforming and holographic projection. In addition, layered semiconductor heterostructures functioning as Schottky diodes or high-electron mobility transistors (HEMTs) benefit from high electron mobility.  Besides, graphene can present advantages over conventional semiconductors, such as remarkable thermal conductivity and outstanding mechanical flexibility, making it highly promising for applications in dynamic wave control at THz frequencies and for realizing the IRS beamforming. Fig. \ref{schematic_illustrations}(a) illustrates a Schottky junction fabricated by integrating metallic split-ring resonators (SRRs) with a thin n-doped gallium arsenide (GaAs) layer. Here, reverse gate-bias voltage dynamically modulates the substrate’s carrier density within the split-gap region, enabling precise resonance tuning of the SRRs.

\subsubsection{Optical Approaches}

Optically driven metasurfaces enable ultrafast modulation of electromagnetic responses through photoconductive effects. In this approach, photosensitive semiconductors, such as silicon, GaAs, and conducting oxides, are illuminated by an external laser beam with photon energy exceeding the material’s bandgap. This photoexcitation generates free carriers, dynamically tuning the material’s conductivity and thereby altering its resonant properties \cite{cong2020spatiotemporal}. A representative implementation is illustrated in Fig. \ref{schematic_illustrations}(b), which features cylindrical resonators fabricated from high-resistivity silicon. When optically excited, these structures achieve subwavelength control over THz waves, with response times as rapid as picoseconds. This ultrafast modulation stems from the near-instantaneous carrier generation and recombination processes in the semiconductor, enabling real-time reconfigurability for applications like high-speed optical switching and adaptive beam steering.

\subsubsection{Phase-change Materials Integration}

Phase-change materials (PCMs), such as vanadium dioxide (VO$_{2}$), chalcogenides (e.g., GeSbTe), and liquid crystals \cite{hou2025220}, empower non-volatile, ultrafast reconfiguration of metasurfaces through reversible structural or electronic phase transitions under external stimuli. These materials dynamically interconvert between amorphous and crystalline states (or distinct crystalline phases) while offering persistent state retention, thermal stability, and ultrafast switching speed. For instance, VO$_{2}$ undergoes a sharp insulator-to-metal transition at $\sim$68°C, switching from a low-conductivity monoclinic phase to a high-conductivity rutile phase. As illustrated in Fig. \ref{schematic_illustrations}(c), integrating VO$_{2}$ into SRRs allows dynamic modulation of the split-gap dimensions via localized phase transitions, thereby tuning electromagnetic resonant responses. Recent advancements \cite{chen2023directional} demonstrated thermally activated THz Janus metasurfaces using VO$_{2}$ meta-atoms, achieving asymmetric transmission and programmable phase delays. In addition, Chalcogenide PCMs, such as GeSbTe, enable analog refractive index tuning through nucleation-mediated crystallinity adjustments, ideal for quasi-continuous optical modulation. These materials can be switched via thermal, electrical, or optical stimuli, allowing for multilevel control and integration into programmable devices.
In contrast, liquid crystals provide broadband electromagnetic control via voltage-induced molecular reorientation.

\subsubsection{MEMS-based Structural Control}

MEMS provide a distinct mechanical approach to metasurface reconfiguration, complementing material-based tuning methods by enabling dynamic electromagnetic control through geometric reconfiguration of meta-atoms. This strategy leverages mature MEMS fabrication technologies to achieve scalable, low-loss THz devices. A representative example, illustrated in Fig. \ref{schematic_illustrations}(d), employs microstructure cantilevers as tunable meta-atoms, where the release angle dynamically tailors electromagnetic responses through mechanical deformation. A novel demonstration \cite{manjappa2018reconfigurable} showcased a reconfigurable MEMS-driven Fano-resonant metasurface with multiple-input-output (MIO) states, where two independent electrical inputs and an optical readout facilitate THz-frequency logic operations via far-field Fano resonance modulation and near-field resonant confinement. Such MEMS-based architectures underscore the potential for scalable, low-loss metasurfaces in THz beamforming.

In summary, reconfigurable metasurfaces are promising for enabling dynamic THz wavefront control in future wireless systems. Electrical tuning modulates material properties via voltage or current (e.g., carrier density in semiconductors, liquid crystal alignment), offering pixel-level precision and nanosecond-scale reconfiguration, though it is limited by wiring complexity. Optical tuning exploits laser-induced effects (e.g., carrier generation, phase-change transitions) for ultrafast, contactless control at the picosecond level, yet it requires external lasers and is prone to thermal interference. Thermal tuning relies on temperature-driven transitions (e.g., VO$_{2}$ phase change) for non-volatile, structurally simple adjustments but suffers from millisecond-scale latency and high energy demands. Hybrid integration of these mechanisms could overcome individual limitations to meet the requirements of advanced terahertz systems. Meanwhile, to expand tunability, hybrid meta-atom strategies, e.g., electromechanical and phase-change integration\cite{prakash2024electromechanically}, enable multifunctional reconfiguration and non-volatile memory effects. These advances solidify metasurfaces as versatile platforms for THz wave engineering in wireless systems.






\section{Main Design Challenges and Solutions}
This section first highlights two critical effects of THz IRS communications, followed by a detailed discussion of  key challenges and potential solutions related to advanced signal processing and  practical deployment. The resulting design insights and corresponding solutions are not limited to IRSs, but readily extend to other reconfigurable metasurfaces, such as active and transmitting metasurfaces.



\begin{figure*}
	\centering
	\includegraphics[width=0.7\textwidth]{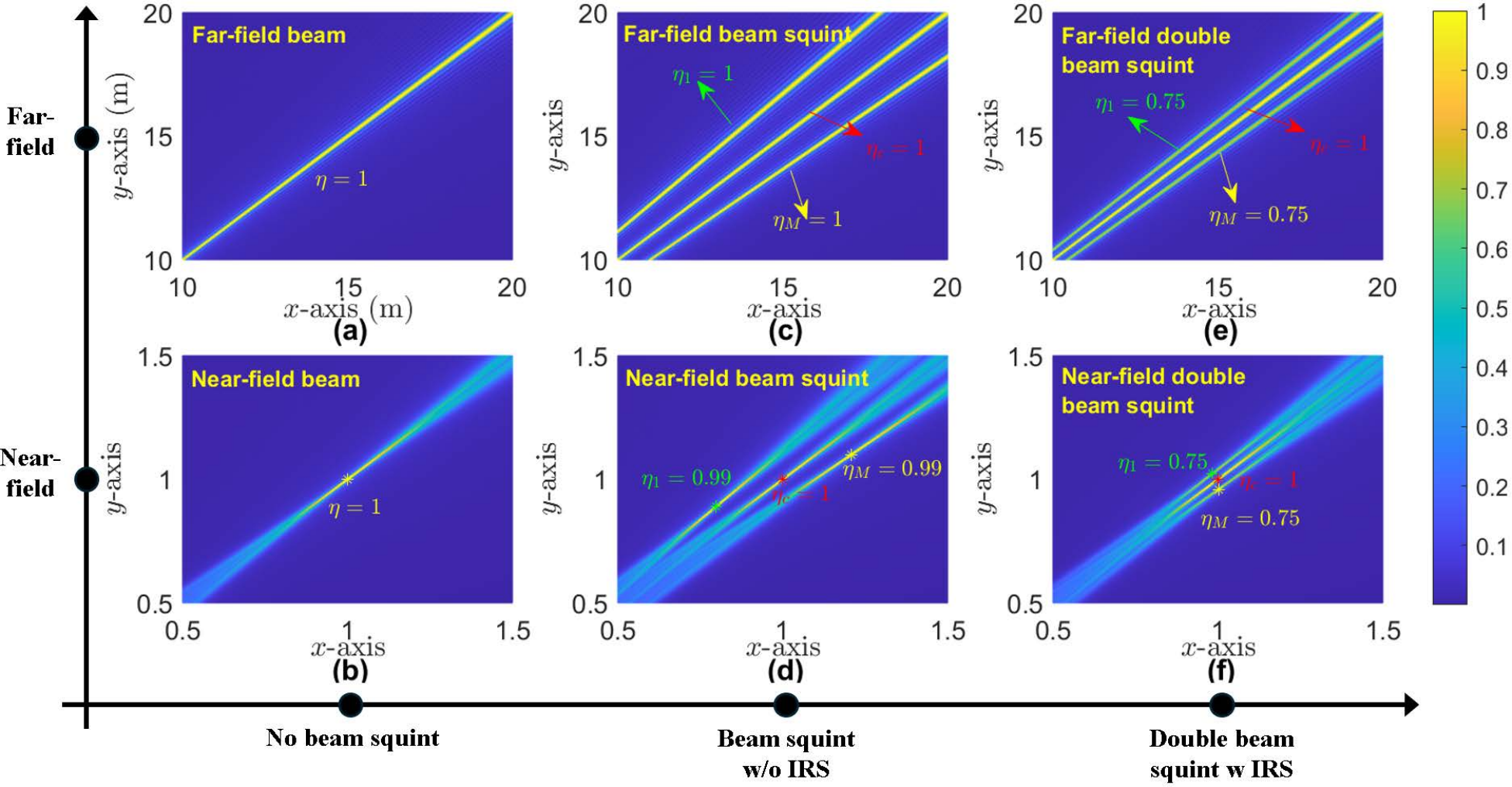}\vspace{-0.3em}\\
	\caption{The beams in the physical space with $\eta_m\in[0,1]$ denoting normalized beam gain at the $m$-th subcarrier: (a) the far-field narrowband scenario; (b) the near-field narrowband scenario; (c) the far-field wideband scenario without (w/o) IRS; (d) the near-field wideband scenario without  IRS; (e) the far-field wideband IRS-assisted scenario exhibiting double beam squint and gain squint; (f) the near-field wideband IRS-assisted scenario exhibiting double beam squint and gain squint. The frequency is 300 GHz and the bandwidth is 30 GHz.}\label{beamsquint}\vspace{-0.3em}
\end{figure*}

\vspace{-3pt}
\subsection{Key Effects: Near Field and Beam Squint Effects}
The integration of IRS to THz communications introduces unique challenges and phenomena that significantly impact signal propagation and system performance. Two critical effects that emerge in this context are the near-field effect and the beam squint effect.

\subsubsection{Near Field Effect}
The near-field range of IRS systems is given by $\frac{r_{1}r_{2}}{r_{1} + r_{2}} < \frac{2D^{2}}{\lambda}$, where $r_{1}$, $r_{2}$, $D$, and $\lambda$ respectively represent the distance between BS and IRS, the distance between IRS and user, the aperture of IRS, and carrier wavelength \cite{Thzmag}. The near-field effect is particularly significant in THz communications due to the electrically larger apertures enabled by shorter wavelengths.
This unique characteristic expands the near-field region of the moderately sized IRS to tens or even hundreds of meters, making it necessary to account for the spherical wavefront propagation. However, accurate near-field channel modeling remains a major challenge, especially under the influence of complex environmental factors such as molecular absorption and scattering, which can vary strongly with frequency, distance, and surrounding conditions. Due to high frequency, even millimeter-scale distance differences can cause noticeable phase shifts, which leads to the high sensitivity of THz signals in user mobility and positioning errors. In addition, the coexistence of users in both near- and far-field regions introduces the so-called near-far field problem, where users experience fundamentally different propagation behaviors, leading to disparities in channel gains and beamforming precision. This poses difficulties in ensuring user quality-of-service  and fairness. Overall, the expanded and complex nature of the near-field in the THz band introduces both new technical challenges and promising opportunities for advanced wireless system design.

\subsubsection{Beam Squint Effect}
In THz band, the beam squint effect arises because the beam directions generated by the base station (BS) vary with frequency. As shown in Fig. \ref{beamsquint}(a), the narrowband far-field beam is transmitted towards a specific direction, where the normalized beam gain is $\eta=1$. However, with the accurate spherical wave model, the narrowband near-field beam is focused on a specific location, as shown in Fig. \ref{beamsquint}(b). As for a wideband system, the severe beam squint effect is induced. In Fig. \ref{beamsquint}(c), the far-field beam squint effect due to the BS is illustrated, which has been extensively discussed in existing literature \cite{su2023wideband}. Although this far-field beam squint causes beams at different frequencies to transmit towards different directions, it does not lead to the beam gain loss across frequencies, as the commonly adopted far-field planar wave assumption ensures that the array gain remains uniform within each beam’s main lobe. However, as aforementioned, such an assumption is not accurate in near-field scenarios. In this case, as shown in Fig. \ref{beamsquint}(d), the near-field beam squint caused by the BS not only directs beams at different frequencies towards different locations, but also leads to beam gain loss at non-center frequencies. Such gain loss is due to minor phase misalignments across the antenna array. Furthermore, with the integration of the IRS, the frequency-independent nature of its phase shifts also introduces beam squint, leading to a more complex double beam squint effect. A key observation in Fig. \ref{beamsquint}(e) is that the reflected beam gain at edge subcarriers is visibly lower than that at the center frequency, while the beams also deviate in direction as this is the far-field case. This degradation originates from the first-stage beam squint at the BS: since the beams at non-center frequencies deviate from the center beam direction, they are no longer well-aligned with the IRS. As a result, only part of their energy is intercepted and reflected by the IRS, i.e., a form of power leakage. A similar pattern is seen with the near-field case in Fig. \ref{beamsquint}(f). We refer to this previously unreported phenomenon as \emph{gain squint}. Note that such gain squint is related to several factors, such as the deployment of IRS, the number of BS antennas, and the operating bandwidth. This observation highlights new challenges for the design of wideband IRS-assisted THz systems, calling for frequency-aware, joint beamforming strategies that account for the spatial misalignment across frequencies.

Near-field effect and the beam squint effect may impact the performance of the THz wireless communication system. In the following, we address four tightly linked challenges that arise from these effects: fundamental performance limit, channel acquisition, beam squint mitigation, and networking of THz IRSs.

\subsection{Fundamental Performance Limit}

This subsection elaborates the performance limit of IRS-aided THz communications to explore essential insights in extreme regimes.

As aforementioned, the severe beam squint effect occurs in wideband system due to frequency-independent components at the BS/IRS that cannot make all the subcarriers achieve their optimal performances. Hence, taking the beam squint effect into account, the scaling law with respect to the number of IRS elements needs to be reconsidered, which is still an open problem. Moreover, the IRS size is usually large in THz systems to compensate for the severe propagation attenuation, which may cause near-field effect. Since the near-field channels have more complex expressions and the IRS is deployed near the BS or users, the BS-IRS/IRS-user channels have the near-field characteristics, which leads to more difficult performance limit analysis. Furthermore, if the IRS size is sufficiently large, both BS and users lie in the near-field region of the IRS such that the above analysis will become further more challenging. However, such problems have not been solved currently. Besides, with the size of BS array and IRS being larger, the beams transmitted from the BS and IRS will be narrower and the beam squint effect will be severer, which will cause the degradation of received power at some subcarriers. This phenomenon points out that the number of effective subcarriers may be remarkably lower than that of transmitted subcarriers, and we call it the squint induced degree of freedom (DoF), which is worth studying. In addition, considering some practical factors of BS and IRS such as the shape, size, antenna/element arrangement, and some novel techniques such as movement and rotation in 3D space, the difficulty of all the above analyses will improve, which needs to be tackled in future works.

For multi-user scenarios, due to the beam squint effect, the beam towards the desired user may lean towards other user’s direction, which leads to more complex multi-user interference. Meanwhile, the hybrid precoding architecture utilized at the BS and the beamforming at the IRS will lead to more complex double beam squint effect, resulting in more challenging capacity regime analysis, which is an open problem at present. Besides, operating at the THz frequency band, the IRS usually has a large size, which leads to significant near-field propagation characteristics and more complex channel modelling. Fortunately, this effect generally leads to high-rank BS-IRS channel in the spatial domain, even for LoS path, which can support multi-stream transmission. However, the schemes for exploiting this merit to improve system performance have not been thoroughly investigated. Furthermore, with an extremely large size of IRS, the high-rank property of BS-IRS channel will hold more probably and the channels between the IRS and different users will be orthogonal. Therefore, the cascaded BS–IRS–user channels may become orthogonal for different users, which merits analytical exploration.

\subsection{Channel Acquisition}
Accurate channel state information (CSI) is vital for IRS-enabled THz beamforming, but remains difficult to acquire due to the passive nature of massive IRS elements. Moreover, near-field effects from large-scale antenna arrays at the BS and IRS further complicate channel acquisition. To tackle these issues, several CSI acquisition methods can be pursued, including conventional channel estimation and beam training-based approaches. Particularly, advanced schemes that exploit the THz beam squint effect have shown promise in enabling faster and more efficient CSI acquisition.

\subsubsection{Conventional Channel Acquisition Schemes} 
In IRS-assisted THz systems, compressive sensing (CS)-based schemes exploit the sparsity of channels in the angular or polar domain to reduce training overhead while maintaining high estimation accuracy. However, their complexity scales with both the number of measurements and the dictionary size. Deep learning-based methods, such as convolutional neural networks and deep belief networks, can capture complex and nonlinear channel characteristics, offering enhanced robustness in dynamic or non-stationary environments, albeit with substantial offline training and high inference complexity, making them more suitable for quasi-static deployments. Kalman filter-based approaches model the channel as a dynamic system, enabling recursive CSI updates based on prior knowledge and observations, making them well-suited for rapidly time-varying THz channels. However, in near-field scenarios, traditional angular-domain sparse estimation techniques become less effective. To address these issues, polar-domain frequency-dependent IRS-assisted channel estimation schemes were proposed, enabling sparse recovery of multipath parameters such as angle, distance, and path gain~\cite{wu2023tcom}. 

Nonetheless, the presence of massive IRS elements introduces a high-dimensional channel matrix, resulting in considerable pilot overhead for conventional estimation methods. To avoid obtaining the full CSI with high dimension, various beam training schemes have been designed, where CSI can be obtained by estimating the physical direction and distance of receiver instead of the entire channel. The most straightforward near-field beam training method is performing a 2D exhaustive search over all possible beam codewords in the angular-distance domains. To reduce the training overhead of the 2D exhaustive search, one efficient approach is a two-stage hierarchical beam training scheme, including the coarse angular estimation stage and the joint angular and distance estimation stage. This approach reduces computational complexity and minimal pilot usage by directly identifying optimal beam pairs, whose accuracy, however, depends heavily on the initial coarse search resolution and may suffer in low SNR conditions. Furthermore, deep learning-aided beam training schemes are promising, in which the best near-field codeword may be predicted by utilizing neural networks, but with extensive offline training. Finally, the BS–IRS and IRS–user links may simultaneously experience both far-field and near-field propagation, resulting in a hybrid-field scenario. This significantly complicates the codebook design process, necessitating the development of novel beam training techniques that are both efficient and accurate.

\subsubsection{Beam Squint-assisted Channel Acquisition  Schemes}

In wideband THz systems, despite its negative impact on beamforming, beam squint creates frequency–angle coupling that can be leveraged for sensing and CSI acquisition. In \cite{gao2023tcom}, a beam squint-assisted channel acquisition scheme has been proposed for THz systems, where beams at different subcarrier frequencies naturally point toward distinct spatial directions. By deploying time-delay devices (TDDs) at the BS, the squint range becomes adjustable, enabling simultaneous sensing of users over a wide angular span. To further expand the sensing and coverage range, the beam split effect arising from inter-antenna spacing larger than half the wavelength is also exploited. Based on this, a joint beam squint and beam split channel acquisition scheme has been proposed for IRS-assisted THz systems \cite{li2025tcom}. In this design, each IRS element is equipped with a TDD, effectively making the IRS frequency-dependent. By jointly optimizing the TDD delays and phase shifts, the resulting beams affected by both squint and split can cover a broad user region. Users then feed back the subcarrier index at which they receive the maximum array gain, allowing the BS to infer their directions and acquire CSI.

In near-field scenarios, beam squint effects become more pronounced, with beam focal points at different subcarriers varying in both direction and distance, as illustrated in Fig. \ref{beamsquint}(b). With the aid of TDDs, the spatial trajectory of beam squint can be manipulated, enabling beams from different subcarriers to distinct target locations. As before, users identify the subcarrier yielding the strongest signal and report its index to the BS, which then estimates their positions or CSI accordingly. Compared to traditional beam steering or scanning-based estimation methods, these beam  squint/split-enabled sensing schemes significantly reduce estimation time and overhead, offering an efficient alternative for CSI acquisition in wideband THz systems. Despite its many benefits, as the beam squint/split becomes more pronounced, the power of each subcarrier beam is dispersed over a wider angular range. While this enhances spatial coverage, it also reduces the beamforming gain at any single direction, which weakens the received signal-to-noise ratio (SNR) and potentially degrades sensing accuracy.


\begin{figure}[t]
	\centering
	\subfloat[]
    {\includegraphics[height=1.3in,width=3.2in]{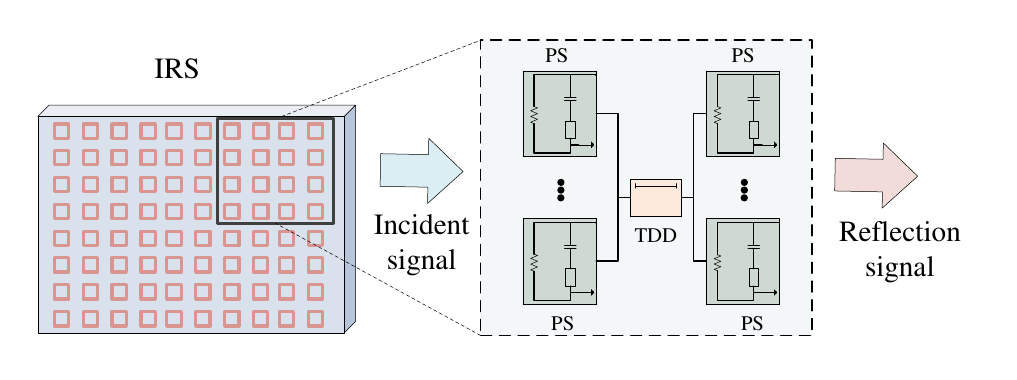}%
		\label{fig_first}}
        \vspace{-10pt}
	\hfil
	\subfloat[]
    {\includegraphics[height=1.35in,width=1.73in]{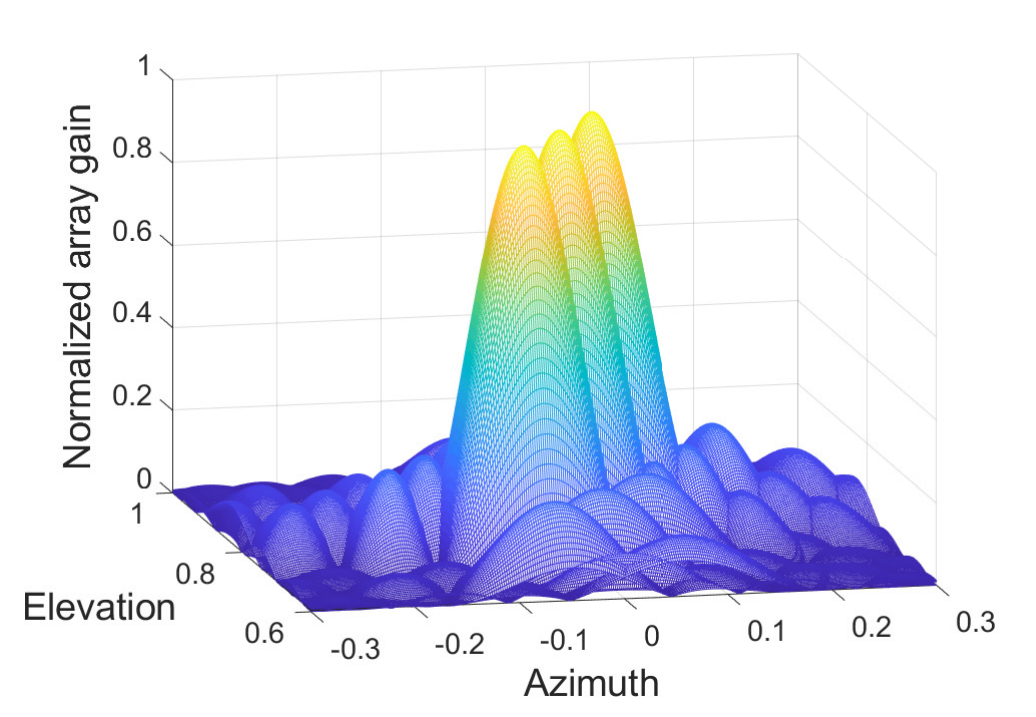}%
		\label{fig_first}}
	\hfil
	\subfloat[]{\includegraphics[height=1.35in,width=1.73in]{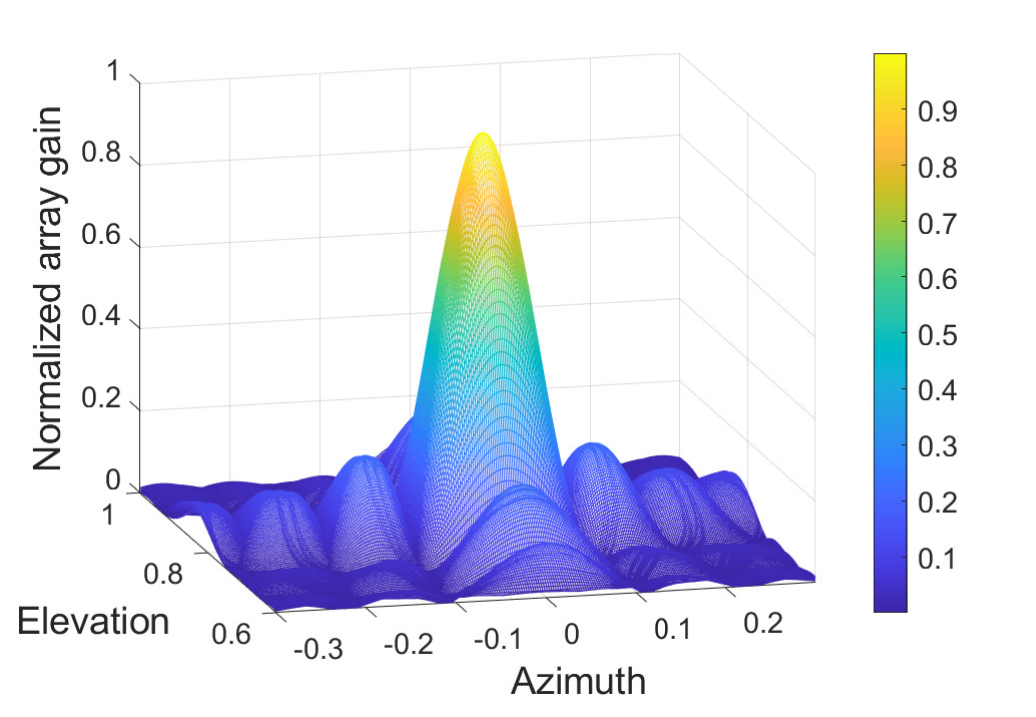}%
		\label{fig_second}}
	\caption{(a) The TDD-based  IRS architecture; (b) Beam pattern under conventional IRS architecture; (c) Beam pattern under TDD-based IRS architecture.}
	\label{fig_beamforming}
\end{figure}

\subsection{Beam Squint Mitigation}

Mitigating the beam squint effect is a significant challenge for achieving accurate beamforming in wideband THz IRS systems. Generally, IRS is a planar structure, and the horizon and elevation directions both generate beam squint effect. 
A simple approach to tackle this issue is to optimize IRS architecture, including its physical shape, size, and so on.  For example, given the same total of IRS elements, a distributed IRS architecture tends to significantly reduce the beam squint effect compared to a centralized IRS architecture, due to smaller array sizes. Moreover,  a square IRS layout is generally more effective than its rectangular counterpart in mitigating this effect. However, this approach, despite being without extra cost, only addresses this effect to a certain extent.  In the following, we discuss advanced strategies to address this issue.

\subsubsection{TDD-based Beam Squint Mitigation} 
THz IRS systems experience a two-stage beam squint effect: from the BS to the IRS and from the IRS to the user. To mitigate the beam squint effect at the BS, one straightforward solution is to replace all phase shifters (PSs) with TDDs. However, this approach incurs excessive power consumption and hardware complexity. To address this issue more efficiently, a hybrid architecture can be adopted where a limited number of TDDs are inserted between the RF chains and the PSs, and thus, the traditional 1D analog beamforming is converted into 2D analog beamforming via the joint control of PSs and TDDs. While introducing TDDs at the BS can alleviate beam squint and enhance system performance, the beam squint issue in the IRS-user link remains unresolved.

A natural solution to address the beam squint at the IRS is to equip each IRS element with a dedicated TDD. However, equipping each IRS element with a TDD is impractical. To strike a balance between performance and implementation feasibility, a TDD-based sub-connected IRS architecture has been proposed \cite{su2023wideband}. In this design, adjacent IRS elements are grouped into subsurfaces, with each group sharing a common TDD module. This significantly reduces the number of TDDs required while retaining sufficient beamforming flexibility. As illustrated in Fig. \ref{fig_beamforming}, the TDD-based sub-connected IRS architecture effectively aligns beams across different subcarriers toward the target direction, mitigating beam squint while maintaining low hardware overhead. This design, although sacrificing some beam-shaping flexibilities, achieves a favorable trade-off between squint mitigation and system complexity, making it well-suited for wideband THz applications. The power consumption of such design depends on the specific hardware interconnection architecture, which can significantly affect efficiency.

\begin{figure}
\centering
\includegraphics[width=0.35\textwidth]{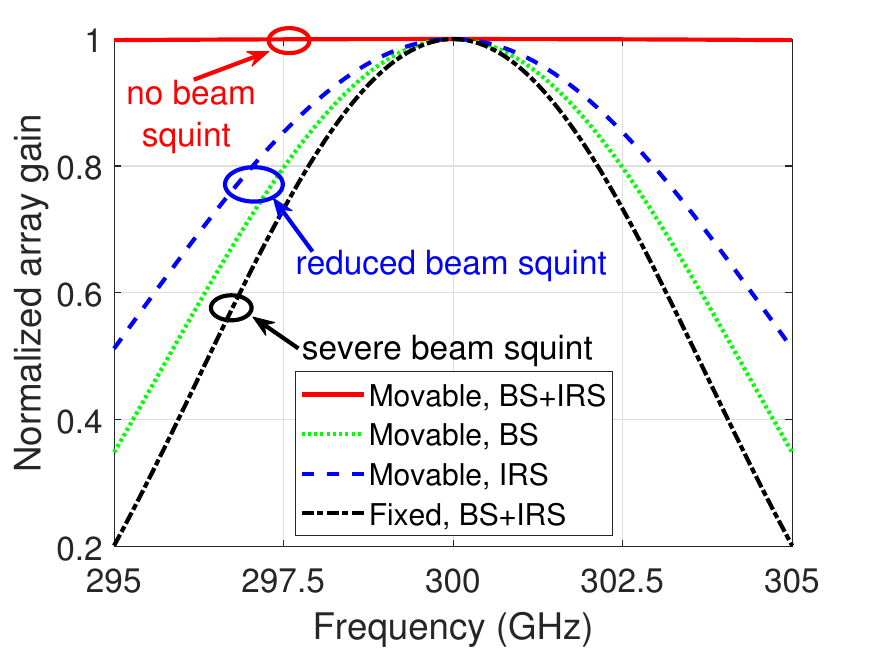}\vspace{-0.3em}\\
\caption{Double beam squint effect mitigation in cascaded
BS–IRS–user link achieved by movable components at both the BS and IRS.}
\label{movable_BS_IRS}\vspace{-0.3em}
\end{figure}

\subsubsection{Movable Array Enabled Beam Squint Mitigation} 
Another novel approach to mitigating the beam squint effect is the movable array technique, where BS/IRS can dynamically configure their antennas/elements' positions to adjust the path delays between them and the receiver's antennas such that beams at different frequencies can be accurately directed toward the intended receiver. Building on this idea, \cite{Zhu2024Suppressing} for the first time successfully eliminated the wideband beam squint effect at the BS by optimizing the positions of movable antennas. Similarly, movable IRS can serve as an effective solution for beam squint mitigation at the IRS side. As shown by Fig. \ref{movable_BS_IRS}, the double beam squint effect previously discussed can be tackled completely by jointly configuring the movable antennas and elements at both the BS and IRS. Note that such an approach is primarily suitable for static or slow-varying channels, while inevitably incurring additional mechanical complexity and energy consumption, making sub-surface reconfiguration a more cost-effective practical alternative. However, whether it can fully mitigate beam squint remains an open question, and achieving an optimal trade-off between performance gain and implementation cost is thus a key direction for future investigation. Beyond 3D position, 3D rotational control of antennas or IRS elements emerges as another promising technique to address beam squint. A natural extension is how to jointly optimize both 3D position and 3D rotation to achieve more precise beam shaping and higher array gain, which is also an open area of research. Finally,  given the mobility-induced cost constraints in practice, it is important to investigate how to allocate the DoF of movable/rotating components across the BS and IRS sides, along with the joint design of the positioning and rotation. This calls for the development of a unified framework to minimize the double beam squint across the entire  cascaded BS–IRS–user link.

In addition to beam squint mitigation in beamforming, efficient resource allocation is also vital for improving system robustness and performance. This may involve a joint design of spatial-temporal-frequency scheduling, BS-IRS-user association, multiple access strategies,  power allocation/control, etc, which requires a general yet efficient optimization framework to tackle such a challenging task given the high-dimensional variables. Furthermore, several practical challenges remain due to hardware impairments in IRS components, such as limited phase resolution, phase noise, mismatch, nonlinear distortion, and mutual coupling effects. These imperfections degrade beamforming accuracy and overall system performance, highlighting the need for advanced compensation techniques to ensure reliable and efficient IRS operation.


\subsection{Networking of THz IRSs}
IRS networking is key to enhancing the capacity and reliability of THz wireless networks. Realizing these gains requires careful design of IRS placement, architecture (passive, active, or hybrid), physical dimensions  (such as size and shape), deployment density, and routing across multi-IRS topologies. Leveraging IRS properties enables networking strategies that balance performance, cost, and complexity.



\subsubsection{Dual Spatial-Scale Deployment} 
Dual spatial-scale IRS deployment includes large-scale and small-/wavelength-scale scenarios, each targeting specific THz communication challenges. 
Large-scale deployment emphasizes placement and architecture to counter distance-dependent THz path loss and expand coverage, generally placing IRSs near the transmitter or receiver \cite{deployment}. Moreover, to handle LoS blockages from buildings, moving objects, and humans, it is promising to integrate 3D environment sensing, light detection and ranging (LiDAR) based mapping, and semantic simultaneous localization and mapping (SLAM) techniques to build real-time visibility maps for IRS placement. In terms of architecture, distributed IRSs generally outperform centralized IRSs in both far- and near-field regimes by increasing channel rank and enabling controllable multipath propagation via spatial diversity, thereby mitigating sparsity-induced rank deficiency at high frequencies. Centralized IRSs offer limited channel rank gains in the far field but can increase channel rank in the near field with large-scale arrays. In addition, increasing the number of IRS elements expands independent signal subspaces in more dimensions, enabling finer beamforming and lower multiuser interference in near-field scenarios. However, it may aggravate beam-split effects, especially under near-field non-linear phase. 

Small-scale or wavelength-scale IRS deployment presents a promising approach to address the challenges of narrow beams and dynamic environments in THz communications. Due to the extremely narrow beamwidth at THz frequencies, even minor positional or angular deviations can cause severe signal degradation. To counter this, cost-effective movable or rotatable IRSs can be employed. In the spatial domain, IRSs can dynamically adjust their 3D position, 3D orientation, or shape to overcome blockages and molecular absorption, enabling precise beam alignment for improved signal strength. In the temporal domain, optimizing movement latency and response time is essential to adapt swiftly to user mobility and environmental changes. Therefore,  the co-design of adaptive control strategies and agile hardware mechanisms is crucial to balance performance gains with mobility-induced overhead.

\subsubsection{Multi-Hop Cooperative Networking} 
Multi-hop IRSs can mitigate THz propagation challenges in cluttered environments by improving coverage and reliability. Most studies select a single reflection path per user, which underutilizes the spatial diversity potential of THz systems. A more general solution is a multi beam multi-hop routing framework under multipath channels, where a multi antenna BS sends orthogonal beams to different IRSs that reflect along distinct paths and the receiver coherently combines them to boost strength and robustness. However, joint path selection and beamforming with machine learning, which warrants further investigation.

Despite its promise, cooperation among multiple IRSs introduces new technical challenges at both physical and network layers. First, cascaded reflection complicates the channel estimation, synchronization, and beam alignment, especially under near-field or wideband conditions. Second, coordinating phase shifts across distributed surfaces rapidly increases the system's control complexity and signaling overhead. Key directions include scalable channel models for cascaded THz IRS channels, graph based algorithms for selecting IRS reflection paths, and active IRS elements to counter cumulative phase errors and attenuation across hops. Finally, artificial intelligence (AI)-based methods offer a promising direction for IRS deployment optimization in scalable and autonomous THz networks. Through learning-based methods that adaptively position and configure decentralized IRSs with environmental awareness, it can effectively overcome key challenges in THz communications, including dynamic channel adaptation, energy-efficient precoding, and multi-hop interference mitigation, while maintaining system scalability.



\begin{figure*}
	\centering
\includegraphics[width=0.88\textwidth]{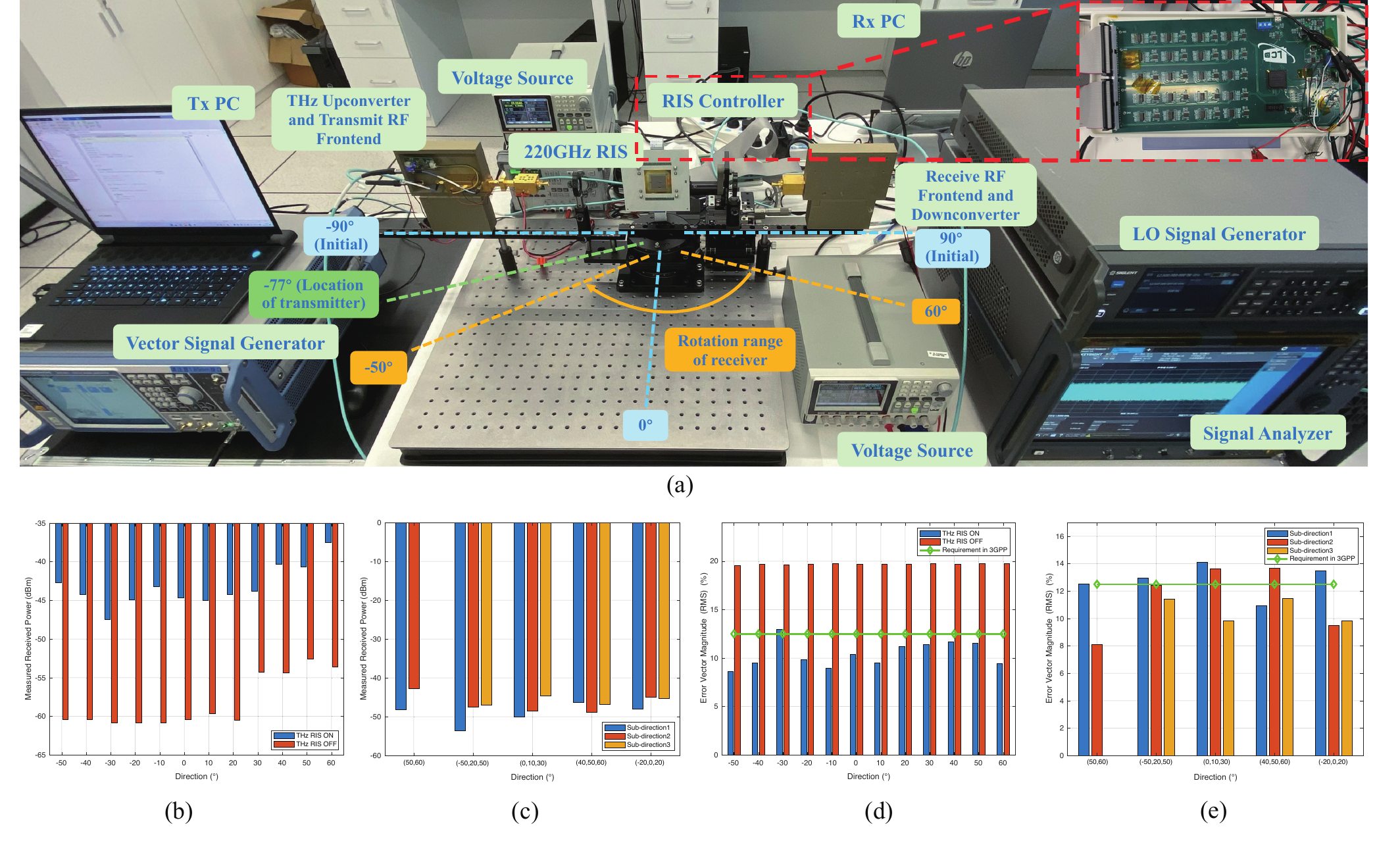}\vspace{-1.0em}\\
	\caption{(a) The prototype of 220 GHz IRS-aided system; (b) measured received power  in the single-user scenario; (c) measured received power in the multi-user scenario; (d)  measured EVM of the received signal in the single-user scenario; (e)  measured EVM of the received signal  in the multi-user scenario. }\label{prototype}\vspace{0.0em}
\end{figure*}

\section{Experiments and Results}




This section presents a prototype of an IRS-aided THz system operating at 220 GHz, as shown in Fig. \ref{prototype}(a), which can be performed over longer distances by increasing antenna or array gain while keeping the power amplifier within its linear region. The liquid‑crystal IRS, which contains 80 strip-shaped units and enables highly directional THz beams in desired directions under a controlled voltage pattern, exhibits a reflection efficiency ranging from $3 \%$ to $16 \%$. The complete system consists of five main modules: baseband transmitter, THz upconversion mixer and transmit RF front-end, IRS, receive RF front-end and THz downconversion mixer, and baseband receiver. The transmitter and receiver are mounted on two rotatable rails, respectively, while the IRS is fixed on the center of the rotation axis.  The angular positions shown in Fig. \ref{prototype}(a) correspond to the initial setup. During the experiment, the transmission link is blocked when the transmitter is positioned at -90°. Experimental results indicate that -77° is an appropriate deployment angle for the transmitter. Meanwhile, the receiver can rotate freely between -50° and 90°. In the single-user scenario, we optimize IRS's beam pattern for 12 directions from -50° to 60° in 10° intervals. In the multi-user scenario, we optimize IRS's sub-beams for five groups of sub-directions including two- and three-user setups: (50°, 60°), (-50°, 20°, 50°), (0°, 10°, 30°),  (40°, 50°, 60°), and (-20°, 0°, 20°), with each angle indicating one user direction. 

Fig. \ref{prototype}(b) shows the measured received power in a single-user scenario across angular directions from –50° to 60°. Compared to the case with the IRS deactivated, the activated IRS provides at least a 15 dB enhancement in received power across all directions. Interestingly, at larger angles such as 40°, 50°, and 60°, relatively higher power levels are observed even without IRS activation. This is attributed to these directions being closer to the specular reflection angle, where sidelobe leakage from the incident signal contributes to increased power.
Fig. \ref{prototype}(c) presents the received power in a multi-user scenario for different sub-beam groups. Compared to the no-IRS case in Fig. \ref{prototype}(b), each sub-beam demonstrates a substantial power gain, highlighting the IRS’s capability in effectively forming directional beams to serve multiple users simultaneously.

Fig. \ref{prototype}(d) illustrates the measured error vector magnitude (EVM) of the received signal across various directions in a single-user scenario, alongside the 3GPP-specified EVM threshold for reliable 16QAM demodulation. When the IRS is deactivated, EVM values in most directions exceed the threshold, indicating poor signal quality and unreliable demodulation. In contrast, activating the IRS significantly reduces the EVM in most directions, bringing them below the threshold and thereby enabling accurate demodulation with lower bit error rates.
Fig. \ref{prototype}(e) shows the EVM performance for different sub-beam groups in a multi-user scenario. Compared to the single-user case in Fig. \ref{prototype}(d), a slight degradation is observed, and certain sub-directions fail to meet the demodulation threshold—suggesting the presence of bit errors. This can be mitigated by deploying a larger IRS aperture in practice. Nevertheless, compared to the IRS-off case, most sub-beams still achieve considerable EVM improvement.
These experimental results validate that integrating IRS into THz communication systems significantly enhances link quality, broadens angular coverage, and supports efficient multi-user transmission.

\section{Conclusions}

This article provided a systematic treatment of IRSs-aided THz communications, covering hardware innovation, signal processing solutions, networking strategies, and experimental validations. We identified that while IRSs offer immense potential in overcoming THz propagation challenges, issues such as near-field effects, beam squint, channel acquisition, and complex deployment scenarios must be carefully addressed. 
Practical results at 220 GHz confirm the feasibility and effectiveness of IRSs in improving signal strength and enabling simultaneous multi-user transmission. Looking forward, further research is needed to develop low-cost, highly tunable IRS elements,  dynamic beam control strategies, and  scalable deployment architectures. Moreover, integrating IRS with sensing, localization, imaging, and resource optimization functionalities will be vital for maximizing the capabilities of future ultra-high-frequency wireless networks, which place more emphasis on feature extraction, data fusion, and classification of the information in contrast to communications focusing on maximizing data throughput and minimizing latency. It is hoped that this work provides a useful resource for the next wave of intelligent and adaptive IRS-enabled THz communications.

The propagation characteristics of THz networks, such as severe path loss, sparse scattering channels, and stringent latency requirements, make the integration of active, hybrid, and omni-directional IRS architectures particularly promising as they collectively address critical challenges through signal amplification, energy-efficient beamforming, and full-space coverage. These advanced IRS architectures enable adaptive and robust THz communications by dynamically balancing performance, power consumption, and multi-user/sensing functionality, representing a vital research direction for next-generation wireless systems.

\bibliographystyle{IEEEtran}
\bibliography{refs.bib}

\end{document}